\documentclass[12pt]{article}
\usepackage{amsmath,amssymb,amsfonts,graphicx}

\begin{document}
\thispagestyle{empty}
\renewcommand{\refname}{References}

\title{\bf Diffraction and quasiclassical limit of \\ the Aharonov--Bohm effect}
\author{Yu.A.~Sitenko$^{1}$ and N.D.~Vlasii$^{1,2}$}
\date{}

\maketitle

\begin{center}
$^{1}$Bogolyubov Institute for Theoretical Physics, \\
National Academy of Sciences, 03680, Kyiv, Ukraine \\
$^{2}$Physics Department, Taras Shevchenko National University of
Kyiv, \\
01601, Kyiv, Ukraine
\end{center}

\begin{abstract}
Since the Aharonov-Bohm effect is the purely quantum effect that has
no analogues in classical physics, its persistence in the
quasiclassical limit seems to be hardly possible. Nevertheless, we
show that the scattering Aharonov-Bohm effect does persist in the
quasiclassical limit owing to the diffraction, i.e. the
Fraunh\"{o}fer diffraction in the case when space outside the
enclosed magnetic flux is Euclidean, and the Fresnel diffraction in
the case when the outer space is conical. Hence, the enclosed
magnetic flux can serve as a gate for the propagation of
short-wavelength, almost classical, particles. In the case of
conical space, this quasiclassical effect which is in principle
detectable depends on the particle spin.
\end{abstract}


\bigskip

\begin{center}
\end{center}

\bigskip
\medskip

\section{Introduction}
The Aharonov-Bohm (AB) effect \cite{Aha} plays a fundamental role in
modern physics. It demonstrates that quantum matter is influenced by
electromagnetic field even in the case when the region of
nonvanishing field strength does not overlap with the region
accessible to quantum matter; the indispensable condition is that
the latter region be non-simply-connected. A particular example is a
magnetic field of an infinitely long solenoid which is shielded and
made impenetrable to quantum matter; such a field configuration may
be denoted as an impenetrable magnetic vortex. Although nowadays the
AB effect is generalized in various aspects and in different areas
of modern physics, in the present letter we shall discuss its
traditional formulation as of a quantum-mechanical scattering effect
off an impenetrable magnetic vortex (see reviews in
\cite{Rui,Ola,Pes,Afa,Ton}). Even in this restricted sense, it
corresponds to two somewhat different but closely related setups.
The first one concerns the fringe shift in the interference pattern
due to two coherent particle beams under the influence of an
impenetrable magnetic vortex placed between the beams. The second
one deals with scattering of a particle beam directly on an
impenetrable magnetic vortex. Almost all experiments are performed
in the first setup, though the second setup is more elaborate from
the theoretical point of view. A direct scattering experiment
involving long-wavelength (slow-moving) particles is hardly
possible, but that involving short-wavelength (fast-moving)
particles is quite feasible, and in the present letter we propose to
perform such an experiment.

In classical theory, scattering off an impenetrable magnetic vortex
is independent of the vortex flux, as well as of the energy of a
scattered particle. Quantum-mechanical scattering depends on the
scattered-particle energy (or wavelength). Although the AB phase
which is acquired by encircling the vortex and related to its flux
is independent of the particle wavelength, the question is how to
observe this phase for different wavelengths and for different
scattering angles. The differential cross section in the limit of
long (as compared to the vortex thickness) wavelengths was shown to
depend periodically on the vortex flux for all scattering angles
\cite{Aha}, and namely this (the periodic dependence on the enclosed
flux) is generically referred to as the AB effect, see, e.g.,
\cite{Ola}. The long-wavelength limit corresponds to the
ultraquantum limit when the wave aspects of matter are exposed to
the maximal extent.

As the particle wavelength decreases, the wave aspects of matter are
suppressed in favour of the corpuscular ones, and therefore the
persistence of the AB effect in the limit of short (as compared to
the vortex thickness) wavelengths seems to be rather questionable.
Actually, there is a controversy in the literature concerning this
point. As it follows from \cite{Rui,Ola}, scattering off an
impenetrable magnetic vortex in the short-wavelength limit tends to
classical scattering which is independent of the vortex flux, and
thus the AB effect is extinct in this limit. On the other hand, it
was already shown by Aharonov and Bohm \cite{Aha} for the case of an
idealized (infinitely thin) vortex that the wave function vanishes
in the strictly forward direction, when the vortex flux equals a
half-of-odd-integer multiple of the London flux quantum; later this
result was generalized to the case of a realistic vortex of finite
thickness \cite{Afa} and, being independent of the value of the
particle wavelength, it persists in the short-wavelength limit.
Thus, this circumstance witnesses in favour of the persistence of
the AB effect, since the wave function for all other values of the
vortex flux is for sure nonvanishing in the forward direction.

The exclusiveness of the forward direction is of no surprise. We
recall the well-known fact that the short-wavelength limit of
quantum-mechanical scattering off a hard core does not converge with
the classical point-particle scattering (perfect reflection), it
differs by a forward peak which is due to the Fraunh\"{o}fer
diffraction; the peak is increasing, as the particle wavelength is
decreasing and the particle is becoming like a classical point
corpuscle. Meanwhile the width of the forward peak is decreasing,
and that is why the experimental detection of the peak is a rather
hard task. As is noted in \cite{Mor}, it seems more likely that the
measurable quantity is the classical cross section, although the
details of this phenomenon depend on the method of measurement. On
the other hand, the forward peak cannot in any way be simply
ignored, because its amplitude is involved in the optical theorem,
whereas the amplitude yielding classical scattering vanishes in the
forward direction. Also, the forward peak contributes considerably
to the total cross section, making the latter twice as large as the
classical total cross section.

Thus, the quantum-mechanical scattering effects persist in the
quasiclassical limit owing to the diffraction effects persisting in
the short-wavelength limit. Concerning the AB effect, this
conjecture is justified quantitatively in the present letter.

However, our consideration is focused mainly on the scattering AB
effect in conical space. Conical space is a space which is locally
flat almost everywhere with exception of a region in the form of an
infinitely long tube; the metric outside the tube is given by
squared length element \cite{Mar, Isr, Sok}
\begin{equation}
{\rm d}s^2=(1-\eta)^{-2}{\rm d}\tilde{r}^2+\tilde{r}^2{\rm d}\varphi^2+{\rm d}z^2={\rm d}r^2+r^2{\rm d}\tilde{\varphi}^2
+{\rm d}z^2,\label{eq1}
\end{equation}
where
$$
\tilde{r}=r(1-\eta),\qquad 0<\varphi<2\pi, \qquad 0<\tilde{\varphi}<2\pi(1-\eta),
$$
and $\eta$ is related to the curvature integrated over the
transverse section of the tube, being of the same sign. Deficit
angle $2\pi\eta$ is bounded from above by $2\pi$ and is unbounded
from below (quantity $-2\pi\eta$ for negative $\eta$ is the proficit
angle that can be arbitrarily large), thus $-\infty<\eta<1$. Conical
space emerges inevitably as an outer space of a topological defect
in the form of a string; such defects known as the
Abrikosov-Nielsen-Olesen vortices \cite{Abr, Nie} arise as a
consequence of phase transitions with spontaneous breakdown of gauge
symmetries, when the first homotopy group of the group space of the
broken symmetry group is nontrivial. Certainly, the value of $\eta$
is vanishingly small for vortices in superconductors, but vortices
under the name of cosmic strings \cite{Kib, Vil} are currently
discussed in cosmology and astrophysics, and the observational data
is consistent with the values of $\eta$ in the range $0<\eta<4\cdot
10^{-7}$ (see, e.g., \cite{Bat}), although the direct evidence for
the existence of cosmic strings is still lacking. In carbon
nanophysics, topological defects in graphene (two-dimensional
crystal of carbon atoms) correspond to nanocones with the values of
$\eta$ equal to positive and negative multiples of $1/6$
\cite{Si7,Si8}. At last, conical space may emerge in a rather
general context of contemporary condensed matter physics which
operates with a variety of two-dimensional structures (thin films)
made of different materials. If such a film is rolled into a cone,
then one can generate quasiparticle excitations in this
conically-shaped film and consider their propagation towards and
through the tip. In all above setups, the problem of
quantum-mechanical scattering of a nonrelativistic particle by a
magnetic vortex in conical space may be relevant.

Scattering in an idealized (with the core of zero transverse size)
conical space was considered by 't Hooft \cite{Ho} and Jackiw et al
\cite{Des,Sou}; later the consideration was extended to the case of
an idealized magnetic vortex placed along the axis of an idealized
conical space \cite{Si2}. However, in the quasiclassical limit the
effects of nonzero transverse size of the core become important.
These effects were taken properly into account in \cite{Si5} (see
also \cite{SiV,S10}), and we shall implicate the results of the
latter works. Instead of using the quasiclassical WKB approximation
or an analogue of the Kirchhoff approximation in optics, we shall
get the quasiclassical limit directly from exact expressions for the
scattering amplitude.

\section{Quantum-mechanical scattering off an impenetrable magnetic vortex}
A scattering wave solution to the Schr\"{o}dinger equation ${\rm
i}\hbar
\partial_t\psi=H\psi$ in the case of a cylindrically symmetric potential has the
following asymptotics at large distances from the symmetry axis:
\begin{equation}
\psi\sim\exp(-{\rm i}Et\hbar^{-1}+{\rm i}k_zz)\left[ \psi_{{\rm in}}({\bf r};\,{\bf k})+
f(k,\,\varphi)\exp({\rm i}kr)\sqrt{\frac {\rm i}r}+O(r^{-1})\right],\label{eq2}
\end{equation}
where ${\bf k}$ is the two-dimensional wave vector which is
orthogonal to the symmetry axis, $k^2=2mE\hbar^{-2}-k_z^2>0$,
$\varphi$ is the angle between vectors ${\bf r}$ and ${\bf k}$, $m$
and $E$ are the mass and the energy of a scattered particle.
Meantime, in the framework of the time-dependent scattering theory,
one gets $S$-matrix:
\begin{equation}
S(k,\,\varphi,k_z;\,k',\,\varphi',k'_z)=\delta(k_z - k'_z)\left[I(k,\,\varphi;\,k',\,\varphi')+\delta(k-k')\frac{\rm i}{\sqrt{2\pi k}}
f(k,\,\varphi-\varphi')\right],\label{eq3}
\end{equation}
where the final ${(\bf k)}$ and initial $({\bf k'})$ two-dimensional
wave vectors are written in polar variables; $f(k,\,\varphi)$ in (2)
and (3) is the scattering amplitude. In the case of the short-range
interaction, the first term in square brackets in (3) is the unity
matrix,
$I(k,\,\varphi;\,k',\,\varphi')=\delta(k-k')k^{-1}\Delta(\varphi-\varphi')$,
where
$\Delta(\varphi)=(2\pi)^{-1}\sum\limits_{n\in\mathbb{Z}}e^{{\rm
i}n\varphi}$ is the delta-function for a compact (angular) variable,
$\Delta(\varphi+2\pi)=\Delta(\varphi)$, $\mathbb{Z}$ is the set of
integer numbers. Appropriately, the incident wave (first term in
square brackets in (2)) is a plane wave, $\psi_{{\rm in}}({\bf
r};\,{\bf k})=\exp({\rm i}kr\,\cos \varphi)$, and the optical
theorem expressing the probability conservation takes form
$2\sqrt{\frac{2\pi}{k}}{\rm Im}f(k,\,0)=\sigma_{\rm tot}$.

If magnetic flux $\Phi$ is enclosed into an impenetrable tube of
radius $r_c$, then the particle wave function obeys the Dirichlet
boundary condition at the edge of the tube, $\psi|_{r=r_c}=0$, and
the Schr\"{o}dinger hamiltonian for a spinless nonrelativictic
particle takes form
\begin{equation}
H=-\frac{\hbar^2}{2m}\left[\partial_r^2+\frac 1r\partial_r+\frac{1}{(1-\eta)^2r^2}\left(\partial_\varphi-
{\rm i}\frac{\Phi}{\Phi_0}\right)^2+\partial_z^2\right],\label{eq4}
\end{equation}
where $\Phi_0=2\pi\hbar ce^{-1}$ is the London flux quantum. The
interaction in this case is not of the potential type and is even
nondecreasing at large distances from the centre (flux enclosed in
the tube).

\subsection{AB effect in Euclidean space}

Already in the case of an impenetrable magnetic vortex in Euclidean
space ($\eta=0$), the long-range nature of interaction leads to a
distortion of the unity matrix in (3), which is now given by
\begin{equation}
I(k,\,\varphi;\,k',\,\varphi')=\cos(\Phi\Phi_0^{-1}\pi)\delta(k-k')k^{-1}\Delta(\varphi-\varphi');\label{eq5}
\end{equation}
appropriately, the incoming wave in (2) is distorted, and factor
$\exp\left[{\rm i}\Phi\Phi_0^{-1}(\varphi-\pi)\right]$ emerges in
addition to the plane wave \cite{Aha}. The scattering amplitude in
this case takes form
\begin{equation}
f(k,\,\varphi)=f_0(k,\,\varphi)+f_{c}(k,\,\varphi),\label{eq6}
\end{equation}
where $f_0$ corresponds to the idealized case of the tube of zero
transverse size (vortex of zero thickness), whereas all the
finite-thickness effects are contained in $f_{c}$. In the
long-wavelength limit, $kr_c\ll 1$, the finite-thickness effects are
negligible, and the scattering amplitude is given by $f_0$ which was
first obtained in \cite{Aha} where it was shown to be a periodic
function of $\Phi$ with the period equal to $\Phi_0$, diverging in
the forward direction. The total cross section diverges as well,
and, although the probability is certainly conserved, the optical
theorem in this limit is hardly informative, being a relation
between two divergent quantities.

On the contrary, in the short-wavelength limit, $kr_c\gg 1$, the
finite-thickness effects are prevailing, since $f_0$ is of order
$\sqrt{r_c}\,O\left[(kr_c)^{-1/2}\right]$, and the scattering
amplitude is given by
\begin{equation}
f_{c}(k,\,\varphi)=f^{({\rm class})}(k,\,\varphi)+f^{({\rm peak})}(k,\,\varphi)+
\sqrt{r_c}\,O\left[(kr_c)^{-1/6}\right],\label{eq7}
\end{equation}
where $f^{({\rm class})}$ yields the classical differential cross
section:
\begin{equation}
\frac{{\rm d}\sigma^{({\rm class})}}{{\rm d}\varphi}=\left|f^{({\rm class})}(k,\,\varphi)\right|^2=
\frac{r_c}{2}\sin \frac{\varphi}{2}\qquad (0<\varphi<2\pi), \label{eq8}
\end{equation}
and $f^{({\rm peak})}$ yields the differential cross section of the
Fraunh\"{o}fer diffraction in the forward ($\varphi=0$) direction:
\begin{eqnarray}
\frac{{\rm d}\sigma^{({\rm peak})}}{{\rm d}\varphi}=\left|f^{({\rm peak})}(k,\,\varphi)\right|^2=
2r_c\left\{\cos(2\Phi\Phi_0^{-1}\pi)\Delta_{kr_c}(\varphi)\right. \nonumber \\
+\left.\left[1-\cos(2\Phi\Phi_0^{-1}\pi)-\sin(2\Phi\Phi_0^{-1}\pi)\sin(kr_c\varphi)\right]
\Delta_{\frac 12kr_c}(\varphi)\right\}.
\label{eq9}
\end{eqnarray}
Here $\Delta_x(\varphi)$ is a regularized (smoothed) delta-function
for the angular variable, which is defined for $-\pi<\varphi<\pi$ by
relations
\begin{equation}
\lim\limits_{x\rightarrow\infty}\Delta_x(\varphi)=\Delta(\varphi),\quad
\Delta_x(0)=\frac{x}{\pi},\label{eq10}
\end{equation}
and, hence,  (9) is strongly peaked in the strictly forward
direction:
\begin{eqnarray}
\frac{{\rm d}\sigma^{({\rm peak})}}{{\rm d}\varphi}=\frac{2}{\pi}kr_c^2\left\{\cos^2(\Phi\Phi_0^{-1}\pi)-
\frac 12kr_c\varphi\sin(2\Phi\Phi_0^{-1}\pi)\right.\nonumber \\\left.-\frac{1}{24}(kr_c\varphi)^2
\left[1+7\cos(2\Phi\Phi_0^{-1}\pi)\right]\right\}\left\{1+O\left[(kr_c\varphi)^2\right]
\right\}, \qquad |\varphi|\ll(kr_c)^{-1}.\label{eq11}
\end{eqnarray}
The optical theorem in the short-wavelength limit takes form
\begin{equation}
\sin^2(\Phi\Phi_0^{-1}\pi)\frac{2\pi}{k}\Delta_{2kr_c}(0)+
2\cos(\Phi\Phi_0^{-1}\pi)\sqrt{\frac{2\pi}{k}}{\rm Im}f^{({\rm peak})}(k,\,0)=\sigma_{\rm tot},\label{eq12}
\end{equation}
where, due to the contribution of the diffraction peak, the total
cross section, $\sigma_{\rm tot}=\sigma^{({\rm
class})}+\sigma^{({\rm peak})}=4r_c$, is twice as large as the
classical total cross section.

\subsection{AB effect in conical space}

If space outside the vortex is conical, then the unity matrix in (3)
is modified in the following way \cite{Si5}:
\begin{eqnarray}
I(k,\,\varphi;\,k',\,\varphi')=\delta(k-k')(2k)^{-1}e^{2{\rm i}k(r_c-\xi_c)}
\left\{\exp\left[{\rm i}\Phi\Phi_0^{-1}(\pi+\omega_\eta)\right]\Delta(\varphi-\varphi'-\omega_\eta)\right. \nonumber \\
\left.+\exp\left[-{\rm i}\Phi\Phi_0^{-1}(\pi+\omega_\eta)\right]\Delta(\varphi-\varphi'+\omega_\eta)\right\},\label{eq13}
\end{eqnarray}
where $\xi_c=\int\limits_{0}^{r_c}{\rm d}s$ is the geodesic radius
of the vortex core (note that spatial region $r<r_c$ is
characterized by nonzero curvature), and
$\omega_\eta=\eta\pi(1-\eta)^{-1}$. The scattering amplitude in the
long-wavelength limit, $f_0(k,\,\varphi)$, is a periodic function of
$\Phi$ with the period equal to $\Phi_0$, diverging in two
directions which are symmetric with respect to the forward one,
$\varphi=\pm \omega_\eta$, see \cite{Sou,Si2,Si5}. These two
directions are the directions along which the Fraunh\"{o}fer
diffraction occurs, and the scattering amplitude in the
short-wavelength limit, $f_{c}(k,\,\varphi)$, is given by
\begin{equation}
f_{c}(k,\,\varphi)=f^{({\rm q-class})}(k,\,\varphi)+f_+^{({\rm peak})}(k,\,\varphi-\omega_\eta)+
f_-^{({\rm peak})}(k,\,\varphi+\omega_\eta)+\sqrt{r_c}\,O\left[(kr_c)^{-1/6}\right],\label{eq14}
\end{equation}
where $f_\pm^{({\rm peak})}$ yields the differential cross section
of the Fraunh\"{o}fer diffraction in one or another direction:
\begin{equation}
\frac{{\rm d}\sigma_\pm^{({\rm peak})}}{{\rm d}\varphi}=
r_c(1-\eta)\Delta_{\frac 12kr_c(1-\eta)}(\varphi\mp \omega_\eta),\label{eq15}
\end{equation}
which, unlike the case of Euclidean space, is independent of the
vortex flux; note that the interference between the amplitudes of
different diffraction peaks is flux dependent, but it is suppressed
as $r_cO\left[(kr_c)^{-1/2}\right]$ in the short-wavelength limit.
Before turning to the remaining part, $f^{({\rm q-class})}$, of the
scattering amplitude in the short-wavelength limit, let us dwell on
classical scattering off a magnetic vortex in conical space.

If the vortex core is impenetrable to a classical point particle,
then its scattering does not depend on the vortex flux and, apart
from the perfect reflection from the core, is purely kinematic. For
a particle with the impact parameter exceeding the core radius,
there is no scattering in coordinates $r,\,\,\tilde{\varphi}$ (see
(1)), but, going over to angular variable $\varphi$, one gets
classical trajectories which after bypassing the vortex either
diverge or converge (and intersect). In the case $-\infty<\eta<0$,
region $\omega_\eta<\varphi<-\omega_\eta$ is not accessible to
incident particles due to the divergence of trajectories; thus, this
region may be denoted as the region of classical shadow. In the case
$0<\eta<1/2$, region $-\omega_\eta<\varphi<\omega_\eta$ is accessed
by incident particles from both sides of the vortex due to the
convergence of trajectories; thus, this region may be denoted as the
region of classical double image \cite{Mar,Vil}. As $\eta$ increases
from 1/2 to 1, the cases of shadow and double image change each
other successively, for more details see \cite{Si5}.

Returning to quantum-mechanical scattering in the short-wavelength
limit, we find that $f^{({\rm q-class})}$ in the case
$-\infty<\eta<0$ is vanishing as
$\sqrt{r_c}\,O\left[(kr_c)^{-1/6}\right]$ in the shadow region,
whereas out of this region it yields the finite differential cross
section which is independent of the vortex flux:
\begin{equation}
\frac{{\rm d}\sigma^{({\rm q-class})}}{{\rm d}\varphi}=\frac 12r_c(1-\eta)^2\sin\left[
\frac 12(1-\eta)\varphi+\frac 12\eta\pi\right]; \label{eq16}
\end{equation}
these results are consistent with classical theory.

In the double-image region in the case $0<\eta<1/2$, $f^{({\rm
q-class})}$ consists of two terms:
\begin{eqnarray}
f^{({\rm q-class})}(k,\,\varphi)=-e^{2{\rm i}k(r_c-\xi_c)}\sqrt{\frac{r_c}{2{\rm i}}}(1-\eta)
\sum\limits_{\pm}\sqrt{\cos\left[\frac 12(1-\eta)(\varphi\mp\pi)\right]}\times \nonumber \\
\times \exp\left\{{\rm i}\Phi\Phi_0^{-1}(\varphi\mp \pi)-2{\rm i}kr_c\cos
\left[\frac 12(1-\eta)(\varphi\mp \pi)\right]\right\};\label{eq17}
\end{eqnarray}
the incident wave in this region consists of two terms as well:
\begin{equation}
\psi_{\rm in}({\bf x};\,{\bf k})=(1-\eta)\sum\limits_{\pm}\exp
\left\{-{\rm i}kr\cos[(1-\eta)(\varphi\mp \pi)]\right\}
\exp\left[{\rm i}\Phi\Phi_0^{-1}(\varphi\mp \pi)\right]. \label{eq18}
\end{equation}
Out of the double-image region, the incident wave is given by one
term, as well as is $f^{({\rm q-class})}$ which yields the
differential cross section in the form of (16). Although the case
$1/2<\eta<1$ can be considered properly, see \cite{Si5}, we shall
restrict ourselves to the case $0<\eta<1/2$ in the following.

Due to the interference between two terms, the differential cross
section corresponding to (17) is dependent on the vortex flux with
the period equal to the London flux quantum:
\begin{eqnarray}
\frac{{\rm d}\sigma^{({\rm q-class})}}{{\rm d}\varphi}=r_{\rm c}(1-\eta)^2\Biggl\{\cos\left(\frac 12(1-\eta)\varphi\right)
\sin\left(\frac 12\eta\pi\right)+ \Biggr.\nonumber \\
\Biggl.+\sqrt{\sin^2\!\left(\frac 12\eta\pi\right)\!-\!\sin^2\!\left(\frac 12(1\!-\!\eta)\varphi\right)}
\cos\!\left[2\Phi\Phi_0^{-1}\pi\!+\!4kr_{\rm c}\sin\!\left(\frac 12(1\!-\!\eta)\varphi\right)\cos\!\left(
\frac 12\eta\pi\right)\right]\Biggr\}, \nonumber \\
|\varphi|<\omega_\eta.\label{eq19}
\end{eqnarray}
In the forward direction in the case
$\sin\left(\frac{1}{2kr_c}\right)\ll\sin\left(\frac12\eta\pi\right)$,
we get
\begin{eqnarray}
\frac{{\rm d}\sigma^{({\rm q-class})}}{{\rm d}\varphi}=2r_{\rm c}(1-\eta)^2\sin(\frac 12\eta\pi)\cos^2
\left[\Phi\Phi_0^{-1}\pi+ kr_c(1-\eta)\varphi\cos\left(\frac 12\eta\pi\right)\right],
\nonumber \\(1-\eta)|\varphi|<(kr_c)^{-1}.\label{eq20}
\end{eqnarray}
The physical reason of the persistence of the AB effect in the
short-wavelength (quasiclassical) limit in the case $0<\eta<1/2$ is
the Fresnel diffraction that is the diffraction in converging rays,
whereas the Fraunh\"{o}fer diffraction is the diffraction in almost
parallel rays.

The dependence on the vortex flux is washed off after integration
over the whole range of the scattering angle,  and we get the total
cross section in the quasiclassical limit,
\begin{equation}
\sigma_{\rm tot}=\sigma^{({\rm q-class})}+\sigma_+^{({\rm peak})}+\sigma_-^{({\rm peak})}=4r_c(1-\eta),\label{eq21}
\end{equation}
which is twice the classical total cross section; the last result is
valid for an arbitrary conical space, $-\infty<\eta<1$. The optical
theorem takes form
\begin{eqnarray}
\cos\left[\Phi\Phi_0^{-1}(\pi+\omega_\eta)\right]\sqrt{\frac{2\pi}{k}}
{\rm Im}\left\{\left[f_+^{({\rm peak})}(k,\,0)+f_-^{({\rm peak})}(k,\,0)\right]e^{-2{\rm i}k(r_c-\xi_c)}\right\}-
\nonumber \\ -\sin\left[\Phi\Phi_0^{-1}(\pi+\omega_\eta)\right]\sqrt{\frac{2\pi}{k}}
{\rm Re}\left\{\left[f_+^{({\rm peak})}(k,\,0)-f_-^{({\rm peak})}(k,\,0)\right]e^{-2{\rm i}k(r_c-\xi_c)}\right\}+
\nonumber \\+ \frac{\pi}{k}\Delta_{2kr_c(1-\eta)}(0) =
\sigma_{\rm tot}.\label{eq22}
\end{eqnarray}

\section{Discussion of results and conclusion}
Thus, we can summarize that the scattering AB effect persists in the
quasiclassical limit owing to the diffraction. Although the effect
is invisible for the cross section integrated over the whole range
of the scattering angle, the effect reveals itself for the
differential cross section. In the case of a magnetic vortex in
Euclidean space, the persistence of the AB effect is due to the
Fraunh\"{o}fer diffraction which is peaked in the forward direction,
see (9) and (11). In the case of a magnetic vortex in conical space,
the peak of the Fraunh\"{o}fer diffraction is shifted from the
forward direction and splitted into two peaks in directions which
are symmetric with respect to the forward one; the contribution of
each peak is independent of the vortex flux, see (15). If the
forward region between two Fraunh\"{o}fer-diffraction peaks is the
region of the classical shadow, then the AB effect disappears in the
quasiclassical limit. If the forward region between two
Fraunh\"{o}fer-diffraction peaks is the region of the classical
double image, then the persistence of the AB effect in the
quasiclassical limit is due to the Fresnel diffraction in this
region, see (19) and (20) for the case $0<\eta<1/2$.

Since a peak of the Fraunh\"{o}fer diffraction is elusive to
experimental measurements, it might be hard to detect the vortex
flux dependence in the strictly forward direction in Euclidean
space. On the contrary, the vortex flux dependence which is due to
the Fresnel diffraction in conical space looks much more likely to
be detectable: it is spread over the wider region in the forward
direction and its amount is finite in the quasiclassical limit,
compare (20) with (11). It should be noted that the optical theorem
in conical space imposes no restrictions on the scattering amplitude
in the strictly forward direction; instead, it involves the
scattering amplitude in the strict directions of the
Fraunh\"{o}fer-diffraction peaks, see (22).

Another distinction of scattering in conical space is that it
depends on spin of a scattered particle: the appropriate spin
connection which is dependent on $\eta$ should be introduced in
hamiltonian (4). In particular, for a spin-1/2 particle the results
are modified in the following way (see \cite{Si5}): one should
change $\Phi\Phi_0^{-1}$ to $\Phi\Phi_0^{-1}\mp \frac 12\eta$, where
two signs correspond to two spin states which are defined by
projections of spin on the vortex axis.

If a magnetic vortex is trapped inside a superconducting
shell\footnote{ Since no magnetic field can leak outside, the
superconducting shell guarantees for certain that there is no
overlap between the region of magnetic flux and the region which is
accessible to the scattered particle, see \cite{Pes,Ton}.}, then its
flux is quantized in the units of a semifluxon, i.e. half of the
London flux quantum. In view of (20) we get the following relation
in the quasiclassical limit when condition
$\sin\left(\frac{1}{2kr_c}\right)\ll\sin\left(\frac
12\eta\pi\right)$ is satisfied:
\begin{equation}
\left.\frac{{\rm d}\sigma^{({\rm q-class})}}{{\rm d}\varphi}\right|_{\Phi=n\Phi_0}+
\left.\frac{{\rm d}\sigma^{({\rm q-class})}}{{\rm d}\varphi}\right|_{\Phi=
\left(n+\frac 12\right)\Phi_0}=2\left.\frac{{\rm d}\sigma^{({\rm class})}}{{\rm d}\varphi}\right|_{\varphi=0},\,\,\,\,
(1-\eta)|\varphi|<(kr_c)^{-1}, \label{eq23}
\end{equation}
where on the right hand-side stands the doubled differential cross
section for the strictly forward scattering of a classical point
particle by a hard core. Hence, the quasiclassical limit of the AB
effect in the case when the vortex flux equals an integer multiple
of a semifluxon can be presented in the form
\begin{equation}
\left.\frac{{\rm d}\sigma^{({\rm q-class})}}{{\rm d}\varphi}\right|_{\Phi=n\Phi_0/2}=\left.F(\varphi,\,\pm)
\frac{{\rm d}\sigma^{({\rm class})}}{{\rm d}\varphi}\right|_{\varphi=0},\,\,\,\,
(1-\eta)|\varphi|<(kr_c)^{-1}, \label{eq24}
\end{equation}
where the upper (lower) sign in $F$ corresponds to even (odd) $n$,
and
\begin{equation}
F(\varphi,\,\pm)=1\pm\cos\left[2kr_c(1-\eta)\varphi\cos\left(\frac 12\eta\pi\right)\right] \label{eq25}
\end{equation}
for a spinless particle,
\begin{equation}
F(\varphi,\,\pm)=1\pm\cos\left[2kr_c(1-\eta)\varphi\cos\left(\frac 12\eta\pi\right)\right]\cos(\eta\pi) \label{eq26}
\end{equation}
for an unpolarized spin-1/2 particle,
\begin{equation}
F(\varphi,\,\pm)=1\pm\cos\left[2kr_c(1-\eta)\varphi\cos\left(\frac 12\eta\pi\right)-\sigma\eta\pi\right] \label{eq27}
\end{equation}
for a polarized spin-1/2 particle ($\sigma=\pm1$ correspond to two
polarization states). We conclude that the enclosed magnetic flux
can serve as a gate for short-wavelength, almost classical,
particles propagating in conical space; the effect depends on the
particle spin, and the most efficient gate is for the propagation of
spinless particles in the strictly forward direction, see (25) at
$\varphi=0$. For instance, the propagation of fast-moving electronic
excitations in the bilayer graphene sample\footnote{ The relevance
of hamiltonian (4) rather than the Dirac hamiltonian is due to the
quadratic dispersion law in this case, see \cite{gra}.} of conical
shape can be governed by the magnetic flux applied through a hole at
the tip.

\section*{Acknowledgments}

The work was partially supported by the Department of Physics and
Astronomy of the National Academy of Sciences of Ukraine under
special program ``Fundamental properties of physical systems in
extremal conditions''.

\end{document}